\documentclass[aps,prd,twocolumn,amsmath,groupedaddress,%
floatfix]{revtex4}
\usepackage{graphicx}

\def\al{\alpha}
\def\be{\beta}

\def\de{\delta}

\def\th{\theta}

\def\la{\lambda}

\def\ta{\tau}

\def\De{\Delta}

\def\fr#1#2{{{#1} \over {#2}}}
\def\frac#1#2{\textstyle{{{#1} \over {#2}}}}
\def\pt#1{\phantom{#1}}

\def\half{{\textstyle{1\over 2}}}
\def\lsim{\mathrel{\rlap{\lower4pt\hbox{\hskip1pt$\sim$}}
    \raise1pt\hbox{$<$}}}
\def\gsim{\mathrel{\rlap{\lower4pt\hbox{\hskip1pt$\sim$}}
    \raise1pt\hbox{$>$}}}

\def\etal {{\it et al.}}

\newcommand{\beq}{\begin{equation}}
\newcommand{\eeq}{\end{equation}}
\newcommand{\bea}{\begin{eqnarray}}
\newcommand{\eea}{\end{eqnarray}}
\newcommand{\bse}{\begin{subequations}}
\newcommand{\ese}{\end{subequations}}
\newcommand{\rf}[1]{(\ref{#1})}

\def\to{\rightarrow}

\def\mix{\leftrightarrow}

\def\nub{\bar\nu}

\def\nh^#1{{\hat N}^{#1}}

\def\hee{h_{ee}}
\def\hmm{h_{\mu \mu}}
\def\htt{h_{\tau \tau}}
\def\hem{h_{e \mu}}
\def\hmt{h_{\mu \tau}}
\def\hte{h_{\tau e}}

\def\tr#1{{\rm{tr}}(#1)}
\def\det#1{{\rm{det}}(#1)}

\def\cLe{(c_L)_{ee}}

\def\ring#1{{\mathaccent'27 #1}}
\def\cri{\ring{c}}
\def\ari{\ring{a}}
\def\mri{\ring{m}}
\def\mrt{{\mri}^2}
\def\mrte{{\mrt}/2E}
\def\mix{\leftrightarrow}

\begin{document}

\title{
Global three-parameter model for neutrino oscillations 
using Lorentz violation}

\author{Teppei Katori}
\author{V.\ Alan Kosteleck\'{y}}
\author{Rex Tayloe}
\affiliation{Physics Department, Indiana University, 
Bloomington, IN 47405, U.S.A.}

\date{IUHET 495, June 2006}

\begin{abstract}
A model of neutrino oscillations is presented 
that has only three degrees of freedom and 
is consistent with existing data.
The model is a subset of the renormalizable sector  
of the Standard-Model Extension (SME),
and it offers an alternative to 
the standard three-neutrino massive model. 
All classes of neutrino data are described,
including solar, reactor, atmospheric, and LSND oscillations. 
The disappearance of solar neutrinos is obtained 
without matter-enhanced oscillations.
Quantitative predictions are offered 
for the ongoing MiniBooNE experiment
and for the future experiments OscSNS, NOvA, and T2K.

\end{abstract}

\maketitle

\section{Introduction}

Local Lorentz invariance is a basic feature
of our best existing theory,
which is the Standard Model (SM) coupled to General Relativity (GR).
This theory is widely believed to be 
the effective low-energy limit
of an underlying structure that unifies quantum physics
and gravity at the Planck scale,
$M_P \simeq 10^{19}$ GeV.
Direct experimentation at this scale is infeasible,
but sensitive measurements might detect 
suppressed low-energy signals of the expected new physics,
including small violations of Lorentz and CPT symmetry
\cite{cpt04,reviews}.
At presently attainable energies,
these unconventional effects can be characterized
in the language of effective field theory 
\cite{kp}.
Extending the GR-coupled SM 
by adding all terms that involve operators for Lorentz violation 
and that are scalars under coordinate transformations
results in a general effective field theory 
called the Standard-Model Extension (SME).
The leading terms in this theory
include those of the SM and GR,
together with ones violating Lorentz and CPT symmetry
constructed from SM and GR fields
\cite{ck,akgrav}.

Neutrinos are natural quantum interferometers,
and so neutrino oscillations represent sensitive phenomena 
with which to conduct searches for physics beyond the minimal SM,
including Lorentz and CPT violation. 
Compelling experimental evidence for oscillations exists,
and they are conventionally attributed to neutrino masses.
In the standard three-neutrino massive model
\cite{pdg},
the effective hamiltonian for neutrino propagation and oscillation
is a 3$\times$3 matrix determined by six parameters,
consisting of two squared-mass differences
$\De m^2_{\odot}$, $\De m^2_{\rm atm}$,
three mixing angles
$\th_{12}$, $\th_{23}$, $\th_{13}$,
and a phase $\de$.
This model can reproduce the observed features 
of the solar-neutrino suppression
\cite{Homestake,SAGE,GNO,Super-Ksol,SNOsol},
the atmospheric-neutrino oscillations
\cite{MACRO,K2K,Super-Kosc,MINOS}, 
and the KamLAND results
\cite{KamLANDosc}.
However,
the model cannot reproduce the LSND signal
\cite{LSND2}.

The presence of Lorentz and CPT violation 
in the underlying theory can produce 
additional or alternative sources of neutrino oscillations
in the low-energy effective Lagrange density,
which are contained in the neutrino sector of the SME.
Even if attention is restricted to three generations
of active neutrinos and dominant effects,
the SME involves many terms and many types of effects
\cite{km1}.
These oscillation effects 
are controlled by coefficients for Lorentz violation
forming various 3$\times$3 matrices 
denoted $(a_L)^\al_{ab}$, $(c_L)^{\al\be}_{ab}$, etc.,
where $\al$, $\be$ are Lorentz indices
and the matrix indices $a$, $b$ label neutrino flavor 
and span $e$, $\mu$, $\ta$.
It is elegant 
to regard these coefficients as arising 
from spontaneous Lorentz violation
\cite{ksp},
which is consistent with Riemann geometry
\cite{akgrav}
and may be ubiquitous in effective field theories
\cite{bak}. 
Some of the neutrino coefficients for Lorentz violation
act to mimic aspects of conventional mass-induced oscillations,
while others predict qualitatively new features of oscillations 
such as 
sidereal variations,
direction dependence,
unconventional spectral behavior, 
Lorentz-violating seesaws,
neutrino-antineutrino mixing,
and more.

The LSND collaboration recently analyzed the sidereal-time dependence
of their reported $\bar \nu_e$ appearance signal
\cite{LVLSND},
using the short-baseline approximation
\cite{km3}. 
Since this signal cannot be incorporated within 
the standard three-neutrino massive model, 
a possible explanation is Lorentz violation. 
The scale required to generate a neutrino-oscillation signal 
at the LSND energies $E\simeq 10^{-2}$ GeV
was found to be of order 
$10^{-19}$ GeV for $(a_L)^\al_{e\mu}$ and $(c_L)^{\al\be}_{e\mu}E$.
This result is compatible with the dimensionless ratio 
$\simeq 10^{-17}$ of electroweak to Planck scales
that might be expected to control
suppression of Planck-scale physics.
The sensitivity achieved is comparable 
to other searches for Lorentz and CPT violation
using interferometric techniques in high-energy physics
involving $K$ oscillations 
\cite{KTeV},
$D$ oscillations 
\cite{FOCUS},
and $B_d$ oscillations 
\cite{BaBar,BELLE,OPAL,DELPHI},
including ones with sidereal variations 
\cite{ak}.
More generally,
the SME sensitivities attainable 
with various types of neutrino searches 
can be comparable 
\cite{km1}
to those achieved in the numerous experiments 
in astrophysics, atomic physics, optical physics, nuclear physics,
and particle physics
\cite{cpt04,reviews}. 

Each coefficient for Lorentz violation leads to 
an energy behavior different from that of a neutrino mass, 
so estimates for the size of possible Lorentz violation 
may be found by considering the energy dependence of the
existing neutrino-oscillation data 
\cite{cg,bpww,bbm,dgpg,hmw}.
With this approach, 
profitable experiments for certain types of coefficients
would involve ultra-high-energy neutrinos
\cite{gghm} 
because the long-baselines and high-energies involved
would yield high sensitivity 
to certain small Lorentz-violating terms.
However, 
it has been shown that some special combinations of 
coefficients for Lorentz violation 
can mimic a mass-like energy dependence 
through a Lorentz-violating seesaw mechanism,
even if the hamiltonian contains no mass terms.
The so-called `bicycle' model
\cite{km2} 
offers a simple example with 
only two coefficients for Lorentz violation.
Despite having no neutrino mass in the hamiltonian,
the model reproduces many features of the
neutrino-oscillation data through a pseudomass term
generated by a seesaw.

In this work, 
we describe a model for neutrino oscillations based on the SME 
that,
like the bicycle model,
uses one CPT-odd coefficient and one CPT-even coefficient. 
However, 
we also introduce a mass term,
so the hamiltonian has three degrees of freedom.
This `tandem' model 
has two Lorentz-violating seesaw mechanisms
that interplay in interesting ways in certain energy regimes.
We demonstrate 
that the model reproduces
the features of all reported neutrino-oscillation data,
including the LSND signal.  
The model and some basic theoretical topics 
are discussed in Sec.\ \ref{tmth},
while the application of the model in the context of 
neutrino-oscillation data
is presented in Sec.\ \ref{application}.
We find that observed $L/E$ behaviors are reproduced
for all baselines $L$ and energies $E$,
and the solar-neutrino result is obtained without
the use of matter-enhanced oscillations
\cite{msw}.  

\section{Tandem model}
\label{tmth}

In this section,
we first discuss some criteria
that are useful in guiding the construction 
of neutrino models in the presence of Lorentz and CPT violation. 
The tandem model is then obtained,
and some of its basic properties are presented.

\subsection{Criteria}
\label{criteria}

In the standard three-neutrino massive model
\cite{pdg},
the effective 3$\times$3 hamiltonian 
describing the propagation and oscillation 
of neutrinos of energy $E$ takes the form
\bea
(h_{\rm SM})_{ab}
& = &
E \de_{ab} 
+ \fr{(m^2_{\rm SM})_{ab}}{2E}.
\label{snm}
\eea
The squared-mass matrix $(m^2_{\rm SM})_{ab}$ 
can be specified by two squared-mass differences,
three mixing angles, and a phase.
Except for the LSND signal,
the current neutrino-oscillation data are consistent
with two large mixing angles and one approximately zero.
It follows that $(m^2_{\rm SM})_{ab}$ 
can be written phenomenologically in the four-parameter matrix form 
\beq
m^2_{\rm SM} \approx
U^T_{\rm SM}
\left(
\begin{array}{ccc}
0 & 0 & 0 \\
0 & \De m^2_{\odot} & 0 \\
0 & 0 & \De m^2_{\rm atm}
\end{array}
\right)
U_{\rm SM}
\eeq
with 
\beq
U_{\rm SM} = 
\left(
\begin{array}{ccc}
1 & 0 & 0 \\
0 & \cos\th_{23} & \sin\th_{23} \\
0 & -\sin\th_{23} & \cos\th_{23}
\end{array}
\right) \hskip-5pt \left(
\begin{array}{ccc}
\cos\th_{12} & \sin\th_{12} & 0 \\
-\sin\th_{12} & \cos\th_{12} & 0 \\
0 & 0 & 1
\end{array}
\right). 
\eeq
The existing data are consistent with 
the parameter values
$\De m^2_{\odot} \simeq8.0 \times 10^{-5}$ eV$^2$, 
$\De m^2_{\rm atm} \simeq 2.5 \times 10^{-3}$ eV$^2$, 
$\th_{12} \simeq 34^{\circ}$, 
and $\th_{23} \simeq 45^{\circ}$.

In this paper, 
we develop an alternative model 
that adequately describes all the existing neutrino-oscillation data,
including the LSND signal.
The effective hamiltonian for neutrino propagation 
in this model contains a term involving conventional neutrino mass,
together with an admixture of Lorentz-violating terms.  
To date,
no compelling evidence for Lorentz violation exists,
so a model of this type is of immediate interest 
only if it is more attractive than the conventional picture
in some other respects.
The standard three-neutrino massive model \rf{snm} 
has a solid foundation in renormalizable quantum field theory,
and it is consistent with all data other than LSND
using only four parameters.
Moreover,
its mass scales $\lsim 0.1$ eV
are compatible with a seesaw origin
\cite{seesaw,seesawreview}. 
Therefore,
if Lorentz violation is to be invoked,
it is desirable to consider models that 
(i) are based on quantum field theory,
(ii) involve only renormalizable terms,
(iii) offer an acceptable description of 
the neutrino-oscillation data,
(iv) have any mass scales $\lsim 0.1$ eV for seesaw compatibility,
(v) involve fewer parameters than the four
used in the standard picture,
(vi) have coefficients for Lorentz violation 
consistent with a Planck-scale suppression $\lsim 10^{-17}$,
and 
(vii) can accommodate the LSND signal.

How challenging is it to satisfy these seven criteria?
To satisfy (i) it suffices to focus on the SME,
since this provides a general field-theoretic framework
for Lorentz violation.
Satisfying (ii) requires restricting attention to 
Lorentz-violating operators of dimension four or less.
However,
this restriction is highly nontrivial 
when taken in conjunction with (iii).
Coefficients for Lorentz violation satisfying (ii) 
generate neutrino oscillations
that are either independent of $E$ or proportional to $E$,
in sharp contrast to the observed behavior proportional to $1/E$
required to satisfy (iii). 

One potential solution to this problem is illustrated 
in the bicycle model
\cite{km2},
which is based on the minimal SME
and describes well the behavior of solar and atmospheric neutrinos 
\cite{mm}
through a Lorentz-violating seesaw mechanism.
The bicycle model has no mass terms
and only two coefficients for Lorentz violation
suppressed by the Planck scale,
so (iv) is irrelevant and (v), (vi) are satisfied.
The observed $E^{-1}$ dependence of oscillations at large energies
emerges as a combination of the Lorentz-violating 
$E^0$ and $E^1$ dependences.
However,
at lower energies the model
predicts a direction-dependent constant-energy signal,
which may be excluded by KamLAND data.
Also,
the LSND signal remains unexplained.

\subsection{Hamiltonian}

The goal of the present work is to provide an explicit
example of a model satisfying all seven criteria (i)-(vii).
To satisfy (i) and (ii),
we adopt Lorentz-violating terms from the minimal SME
but omit for simplicity 
the provision for neutrino-antineutrino mixing. 
In this context, 
the effective hamiltonian 
for neutrino propagation takes the form 
\cite{km1}
\bea
(h^\nu_{\rm{eff}})_{ab}
& = &
E{\de}_{ab}+\fr{(m^2)_{ab}}{2E}
\nonumber\\
&&
\qquad +\fr{1}{E}
[(a_L)^{\mu} p_{\mu} - (c_L)^{\mu \nu} p_{\mu} p_{\nu} ]_{ab}.
\label{mSME}
\eea
Since the coefficients $(a_L)^{\mu}_{ab}$
are associated with CPT-odd operators
in the Lagrange density, 
the effective hamiltonian 
$(h^{\bar \nu}_{\rm{eff}})_{ab}$
for antineutrino propagation
is obtained by reversing the sign of the coefficients 
$(a_L)^{\mu}_{ab}$.
The coefficients $(a_L)^{\mu}_{ab}$ 
have dimensions of mass,
while $(c_L)^{\mu\nu}_{ab}$ are dimensionless.
In obtaining Eq.\ \rf{mSME},
possible gravitational couplings 
\cite{akgrav,baik}
have been disregarded,
and the coefficients 
$(a_L)^{\mu}_{ab}$ and $(c_L)^{\mu\nu}_{ab}$ 
are assumed to be spacetime constants.
If these coefficients originate in spontaneous Lorentz violation,
they have companion Nambu-Goldstone fluctuations 
that could be interpreted as 
the photon \cite{bkng},
the graviton \cite{kpng},
or additional neutrino interactions
\cite{gktw},
but these effects are secondary in the present context
and are disregarded in this work.

The occurrence of the neutrino three-momentum $\vec p$
in the hamiltonian \rf{mSME}
means that the oscillation physics in the chosen inertial frame
typically depends on the direction of neutrino propagation 
\cite{km1}.
The number of degrees of freedom can be significantly reduced
if this complication is avoided.
One possibility is to suppose the model is rotationally invariant.
In a Lorentz-violating theory,
this requirement can be implemented 
only in a single special inertial frame.
If this frame is identified
with that of the cosmic microwave background radiation,
for example,
then direction-dependent effects 
are still present in neutrino-oscillation experiments
because the solar system is moving with respect to this frame.
However,
the rotational, orbital, and translational motions of the Earth
are essentially nonrelativistic
in the approximately inertial Sun-centered frame
\cite{sunframe}
relevant for neutrino experiments,
so in practice any direction-dependent effects
in models of this type are suppressed to parts in a thousand or more.
Another possibility is to suppose the model
does have large direction-dependent effects
but that they are irrelevant for describing
the available experimental data.
The point is that reported neutrino-oscillation results 
are typically obtained by integrating data 
taken over long time periods,
so the average values of the direction-dependent coefficients
may suffice even in theories 
with large direction-dependent effects.
In what follows, 
we disregard direction-dependent effects
as a reasonable first approximation,
without precluding their existence
in a more detailed treatment.

With this assumption,
the neutrino effective hamiltonian can be written in the form
\bea
(h^\nu_{\rm{eff}})_{ab}
& \approx &
E{\de}_{ab}+\fr{(m^2)_{ab}}{2E}
+(a_L)_{ab} -\frac 4 3 (c_L)_{ab}E.
\label{isoham}
\eea
In practice,
terms proportional to the unit matrix 
can be disregarded because they produce no oscillation effects.
Note, however,
that mass terms proportional to the unit matrix
may play a role in ensuring stability and causality
of the underlying theory
\cite{kleh}. 

To maintain the possibility of satisfying criteria (iii) and (vii)
while keeping the number of degrees of freedom small,
we further restrict attention to effective hamiltonians
of the form \rf{isoham}
adapted from the simple scheme of the bicycle model
\cite{km2}.
In that model,
the high-energy pseudomass is created from
a Lorentz-violating seesaw mechanism
involving a $(c_L)$-type coefficient 
in the on-diagonal $(h_{\rm eff})_{ee}$ component
together with off-diagonal $(a_L)$-type coefficients.
To incorporate also the observed KamLAND $L/E$ dependence
\cite{KamLANDosc}
and generate effects that can reproduce the LSND signal
\cite{LSND2},
a second `tandem' Lorentz-violating seesaw can be introduced 
that operates at low energies.
This can be triggered by adding another on-diagonal entry
that is located in the $(h_{\rm eff})_{\ta\ta}$ component
and involves an $(m^2)$-type parameter.
These considerations suggest limiting attention
to the special case of the hamiltonian \rf{isoham}
taking the form
\bea
h^\nu _{\rm eff}&=&
\left(
\begin{array}{ccc}
-\frac 4 3(c_L)_{ee}E & ~ (a_L)_{e\mu} & ~ (a_L)_{e\tau} \\
(a_L)_{\mu e} & ~0 & ~ (a_L)_{\mu\tau} \\
(a_L)_{\tau e} & ~ (a_L)_{\tau\mu} & ~(m^2)_{\ta\ta}/2E
\end{array}
\right).
\label{fivedof}
\eea
Since this hamiltonian is hermitian and the elements are real,
it is symmetric.
The model \rf{fivedof} therefore has five degrees of freedom.

To satisfy criterion (v), 
we must reduce the degrees of freedom to fewer than four.
We find that the current data may be described 
with the simplifying assumptions that 
all the $(a_L)$-type coefficients are identical, 
$(a_L)_{e\mu}=(a_L)_{\mu \tau}=(a_L)_{\tau e}$. 
For simplicity,
following the suggestive notation of Ref.\ \cite{km1},
we write $m_{\ta\ta}=\mri$, $a_L = \ari$, and $-4\cLe/3=\cri$. 
The effective hamiltonian for neutrinos becomes 
\bea
h^\nu_{\rm TM}&=&
\left(
\begin{array}{ccc}
\cri E & ~ \ari & ~ \ari \\
\ari &  ~0 & ~ \ari \\
\ari & ~ \ari & ~\mrte
\end{array}
\right).
\label{tandemh}
\eea
This is the tandem model,
which depends on only three independent degrees of freedom
and is the focus of the remainder of this work.
Note that the presence of CPT violation implies that
the corresponding effective hamiltonian
for antineutrinos is 
\bea
h^{\bar \nu}_{\rm TM}&=&
\left(
\begin{array}{ccc}
\cri E & ~ -\ari & ~ -\ari \\
-\ari &  ~0 & ~ -\ari \\
-\ari & ~ -\ari & ~\mrte
\end{array}
\right).
\label{antitandemh}
\eea

\begin{table*}
\renewcommand{\arraystretch}{1.5}
\begin{tabular}{lccc}
\hline
\hline
Feature & Standard model \cite{pdg}
& \pt{x} Bicycle \cite{km2} \pt{x} & \pt{x} Tandem \pt{x} \\ 
\hline
Lagrange-density formulation compatible with other physics 
& yes & yes & yes \\
renormalizable terms only
& yes & yes & yes \\
generations of active neutrino species
& 3 & 3 & 3 \\
number of sterile neutrinos 
& 0 & 0 & 0 \\
degrees of freedom in effective hamiltonian 
& 4 to 6$^a$ & 2 & 3 \\
global description of neutrino-oscillation data except LSND
& yes & yes$^b$& yes \\
description of LSND result 
& no & no & yes \\
matter-enhanced oscillations needed 
& yes & yes & no \\
seesaw-natural masses (if any) $\lsim$ 0.1 eV
& yes & none & yes \\
coefficients for Lorentz violation (if any) $\lsim 10^{-17}$ suppressed
& none & yes & yes \\
CPT-odd operators 
& no & yes & yes \\
large direction-dependent effects 
& no & yes & no$^c$ \\
neutrino-antineutrino mixing$^d$
& no & no & no \\
\hline
\hline
\end{tabular}
\begin{flushleft}
\vskip -10 pt
\begin{tabular}{l}
\quad \qquad
$^a${\footnotesize 
Depending on whether $\th_{13}$ and $\de$ are included.
}
\end{tabular}
\\
\vskip -5 pt
\begin{tabular}{l}
\quad \qquad
$^b${\footnotesize 
Except perhaps KamLAND (cf.\ Sec.\ \ref{criteria}).
}
\end{tabular}
\\
\vskip -5 pt
\begin{tabular}{l}
\quad \qquad
$^c${\footnotesize 
In the isotropic approximation.
}
\end{tabular}
\\
\vskip -5 pt
\begin{tabular}{l}
\quad \qquad
$^d${\footnotesize 
See Ref.\ \cite{km1} for models with this feature.
}
\end{tabular}
\end{flushleft}
\vskip -10pt
\caption{
Some attributes of the standard three-neutrino massive model,
the bicycle model, and the tandem model.
}
\label{features}
\end{table*}

In Sec.\ \ref{application},
we show that a choice for the three degrees of freedom 
can be made that respects criteria (iv) and (v)
and that yields a global description 
of the current neutrino-oscillation data
including the LSND signal.
This means that the tandem model satisfies
all seven criteria (i)-(vii),
making it a useful alternative candidate model
for neutrino oscillations.
In fact,
the structure of neutrino oscillations
in the tandem model is remarkably rich
despite the presence of only three degrees of freedom.
This is a consequence of the double Lorentz-violating seesaw
and the accompanying CPT violation,
and it predicts a variety of observable phenomena 
in future experiments. 
Table I provides a summary comparison of some attributes of 
the standard three-neutrino massive model,
the bicycle model,
and the tandem model.

\subsection{Properties}

To analyse neutrino mixing in the tandem model,
we diagonalize the hamiltonian \rf{tandemh}
using a 3$\times$3 unitary mixing matrix $U$,
\beq
h^\nu_{\rm TM} 
= U^\dagger E_{\rm TM} U,
\label{tmdiag}
\eeq
where $E_{\rm TM}$ is a $3\times3$ diagonal matrix
containing the $h^\nu_{\rm TM}$ eigenvalues $E_J$, $J = 1,2,3$. 
The description of neutrino oscillations then depends
on the mixing matrix elements $U_{Ja}$ 
and the eigenvalue differences $\De_{JK}=E_J-E_K$.  

Since $h^\nu_{\rm TM}$ is nondiagonal and includes terms 
with distinct dependences on the neutrino energy $E$, 
both $U_{Ja}$ and $\De_{JK}$ 
have intricate energy behavior. 
Moreover,
the $E$ dependences of the corresponding quantities 
for the antineutrino hamiltonian \rf{antitandemh}
are different.
As a result, 
the oscillation probabilities for neutrinos and antineutrinos 
can vary strongly and distinctly with $E$.
For example, 
we show in the next section that
the dependence of $U_{Ja}$ on $E$ 
results in energy variations 
for solar pp- and $^8$B-neutrino oscillations
analogous to those arising via matter-enhanced effects
\cite{msw}
in the standard three-neutrino massive model.
Similarly,
the dependence of $\De_{JK}$ with $E$
produces behavior matching the data 
for both the KamLAND reactor 
and the Super-Kamiokande atmospheric neutrinos. 
Also,
an oscillation signal for LSND emerges.

To perform the diagonalization explicitly,
we used two different procedures:
numerical matrix diagonalization,
and analytical solution of the cubic eigenvalue equation.
The two procedures yield the same results. 
For the numerical work, 
we adopted the simple Jacobi method \cite{ppvf}.
An appropriate orthogonal matrix (Jacobi rotation)
is created and applied to the hamiltonian
to eliminate one off-diagonal component.
This is repeated in turn for the other two off-diagonal elements,
thereby completing one Jacobi sweep.
Five sweeps yield diagonalization at sufficient precision,
permitting calculation of the eigenvalue differences $\De_{JK}$ 
and mixing-matrix elements $U_{Ja}$.
For the analytical method, 
the standard del Ferro-Cardano solution 
of the cubic equation was used to obtain the three eigenvalues
and the corresponding eigenvectors 
and to construct the mixing matrix. 
Some details of this solution are given in the Appendix. 
In practice, 
the diagonalization procedure was repeated for
each value of the neutrino energy $E$ of interest. 

The neutrino-oscillation probability in the SME framework is derived
in Ref.\ \cite{km1}.  
For the tandem model,
all coefficients in $h^\nu_{\rm TM}$ are real
and the oscillation probability for neutrinos 
reduces to the simple form
\beq
P_{\nu_a\to\nu_b}=\de_{ab}-4\sum_{J>K}
U_{Ja}U_{Jb}U_{Ka}U_{Kb}\sin^2 {\left(\half{\De_{JK} L}\right)},
\label{tmosc}
\eeq
where $L$ is the baseline distance. 
The oscillation probability $P_{\nub_b\to\nub_a}$
for antineutrinos is given by an expression
of the same form,
but with $U_{Ja}$ and $\De_{JK}$ 
obtained by diagonalizing $h^{\bar \nu}_{\rm TM}$ instead.

Since the effective hamiltonian $h^\nu_{\rm TM }$ is symmetric,
$P_{\nu_a\to\nu_b} = P_{\nu_b\to\nu_a}$,
as may also be seen from Eq.\ \rf{tmosc}.  
This is a property associated with T invariance.
In fact,
the Lorentz violation arising via 
nonzero isotropic coefficients $\ari$ and $\cri$ preserves T symmetry.
The presence of the coefficient $\ari$ 
implies CPT violation,
but detecting this violation 
via $P_{\nu_a\to\nu_b} \neq P_{\nub_b\to\nub_a}$ 
may be challenging in certain energy regimes,
as discussed in the next section. 
Since the tandem model is CPT violating but T invariant,
it breaks CP symmetry.
More generally,
Lorentz violation maintaining rotation invariance
and controlled by coefficients of the $a_L$ and $c_L$ types
preserves T symmetry
but violates CP through a combination of C and P breaking 
\cite{km1,klp}.
Note that this origin of CP violation
is qualitatively different from that 
in the standard three-neutrino massive model,
which arises via a phase in the mixing matrix $U$
\cite{pdg}.

Finally,
we offer a few remarks about model building. 
At present,
no completely satisfactory theory of lepton and quark masses
is available.
In the standard three-neutrino massive model,
understanding the neutrino mass matrix
offers some unique challenges
associated with the origin, values, and stability
of the masses and mixing angles. 
The standard picture provides partial answers
by invoking one or more seesaw mechanisms
in a grand-unified theory, 
perhaps with supersymmetry
\cite{ms,nubooks}.
In this context,
provided criterion (iv) is satisified
in the selection of the value of $\mrt$,
the tandem model is on a roughly comparable footing
to the standard three-neutrino massive model.
However,
the required structure of the neutrino mass matrix 
is somewhat different,
which offers interesting scope for model building. 
The essential issue for the tandem model is to explain
the dominance of the component $(m^2)_{\ta\ta}$
in the mass matrix in Eq.\ \rf{mSME}.
This can arise naturally in some cases.
For example,
in a simple SO(10) grand-unified theory
the neutrino Dirac-mass matrix can be proportional to
the quark mass matrix
\cite{nubooks,fyy},
so invoking a seesaw mechanism
can produce neutrino masses proportional to the square
of the quark masses.
Under these circumstances,
the large mass of the $t$ quark
can ensure that the neutrino mass matrix
is dominated by $(m^2)_{\ta\ta}$,
as is effectively assumed in the tandem model.
Analogous model-building issues exist 
for the coefficients for Lorentz violation $\cri$ and $\ari$
in the tandem model,
and these represent an interesting open area for future work.

\section{Application}
\label{application}

We find that the tandem model 
provides a good description of existing neutrino-oscillation data 
with the following values for the mass parameter and the
two coefficients for Lorentz violation 
in Eqs.\ \rf{tandemh} and \rf{antitandemh}:
\bea
\half \mrt & = & 5.2\times 10^{-3} ~ \mathrm{eV}^2,
\nonumber\\
\ari & = & - 2.4 \times 10^{-19} ~ \mathrm{GeV} ,
\nonumber\\
\cri & = & 3.4 \times 10^{-17} .
\label{values}
\eea
Note that this choice of $\mri$ respects the cosmological constraint 
on neutrino masses
\cite{mt}.
Also,
the values for $\ari$ and $\cri$ are consistent 
with the results extracted from LSND data 
\cite{LVLSND,km3}.

The choice of values for $\mrt$, $\ari$, and $\cri$
is guided by the presence of the double Lorentz-violating seesaw
in the tandem model.
Taken independently,
each seesaw generates distinct asymptotic effects,
but by continuity the two must merge in some energy regime.
Near the merger scale, 
the neutrino behavior is controlled by the interplay of both seesaws,
and interesting effects appear.
The values \rf{values} are chosen to
yield these merger effects around 10 MeV,
while still satisfying criteria (iv) and (vi).
It is possible that other values for $\mri$, $\ari$, and $\cri$ exist 
that are compatible with the neutrino-oscillation data.

In this section,
we begin by describing some general features 
predicted by the model with the values \rf{values}.
We then offer a comparison to existing results 
from a variety of neutrino-oscillation experiments.

\subsection{Features}

\begin{figure}
\centering
\includegraphics[width=\columnwidth]{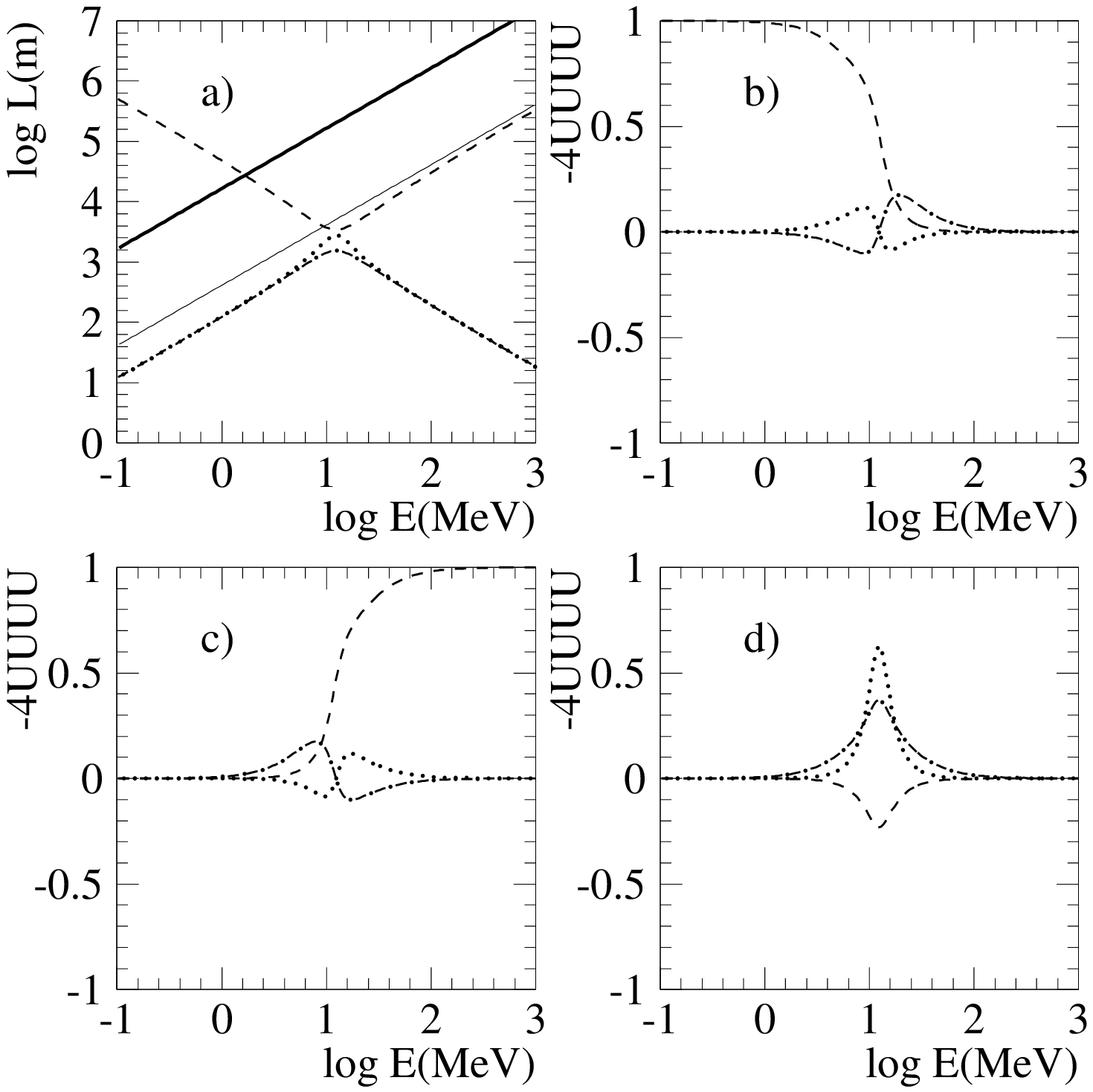}
\caption{
Neutrino behavior in the tandem model.
(a) Location of the first oscillation maximum 
in the $\log L$-$\log E$ plane.
The dashed, dotted, and dash-dotted lines correspond
to $\De_{21}$, $\De_{32}$, and $\De_{31}$, 
respectively.
For comparison,
the thin and thick solid lines show the location 
for $\De m^2_{\rm atm}$ and $\De m^2_{\odot}$,
respectively,
in the standard three-neutrino massive model.
(b) Values of $-4U_{Ja}U_{Jb}U_{Ka}U_{Kb}$
versus energy for $\nu_e\mix\nu_\mu$ oscillations. 
The dashed, dotted, and dash-dotted lines correspond
to $(J,K) = (2,1)$, $(3,2)$, and $(3,1)$, 
respectively.
(c) Same for $\nu_\mu\mix\nu_\ta$ oscillations.
(d) Same for $\nu_e\mix\nu_\ta$ oscillations.
}
\label{fig:nusol}
\bigskip
\centering
\includegraphics[width=\columnwidth]{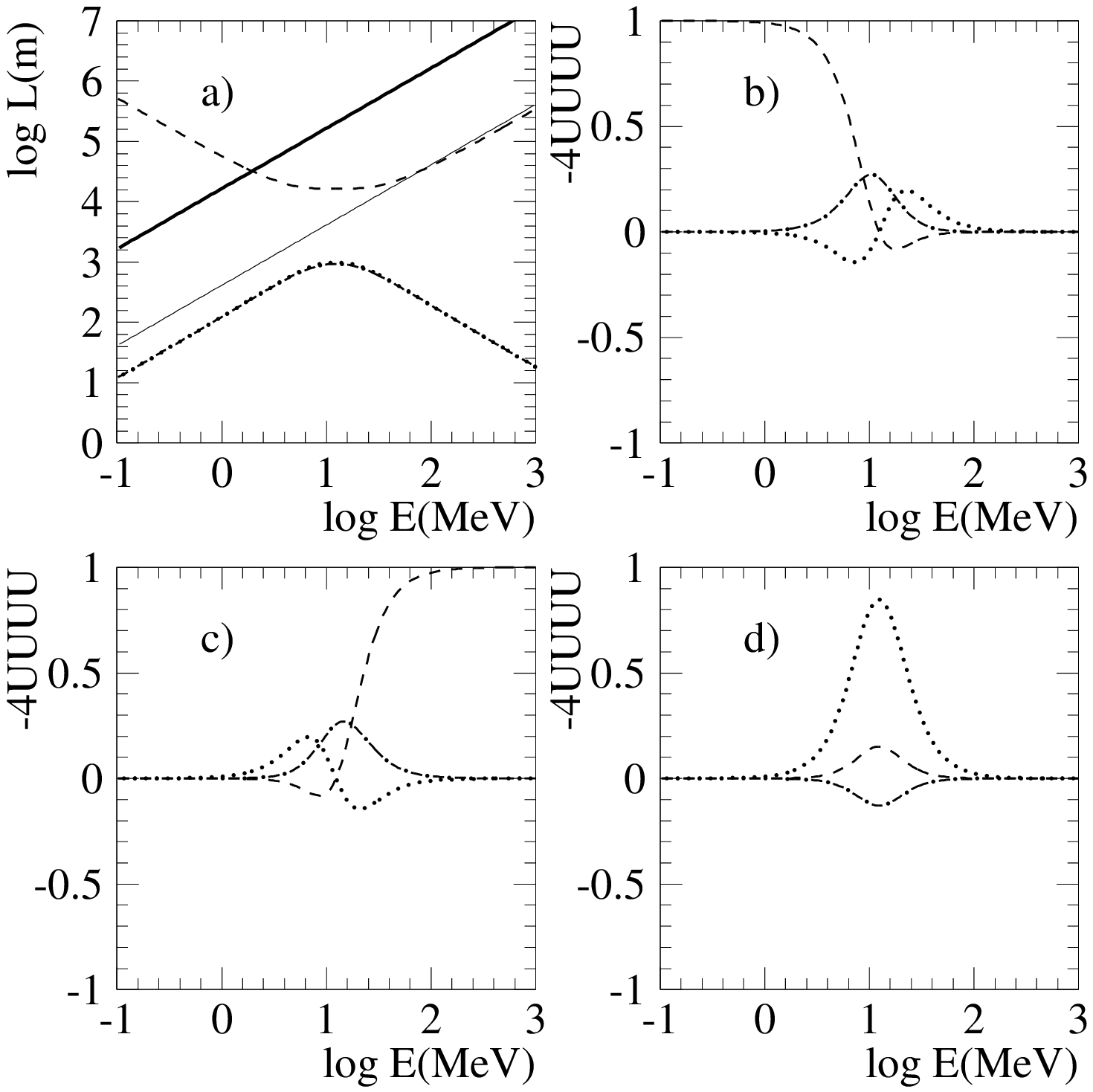}
\caption{
Antineutrino behavior in the tandem model.
The descriptions of (a)-(d) match those of 
Fig.\ \ref{fig:nusol}.  
In (a), the dotted and dash-dotted lines appear to coincide
but in fact remain distinct for all $E$,
as explained in the Appendix. 
}
\label{fig:antinusol}
\end{figure}

Some general features of the tandem model 
with the choices \rf{values} 
are shown as a function of energy
for neutrinos in Fig.\ \ref{fig:nusol} 
and for antineutrinos in Fig.\ \ref{fig:antinusol}.
These figures reveal a relatively complicated energy dependence,
arising because the energy dependences
of the eigenvalue differences $\De_{JK}$
and the elements of the mixing matrix $U_{Ja}$
are more involved than the simple power law
of the standard three-neutrino massive model.
As can be seen by comparing 
Figs.\ \ref{fig:nusol} and \ref{fig:antinusol},
the oscillation lengths and probabilities 
of neutrinos and antineutrinos differ in detail.
This occurs because the coefficient $\ari$ is 
associated with CPT violation 
and enters with the opposite sign in the two cases. 

Figures \ref{fig:nusol}(a) and \ref{fig:antinusol}(a) 
display curves in the $\log L$-$\log E$ plane 
that establish experimental sensitivities
to oscillations involving particular flavors.
The oscillation length $L$ for each type of oscillation 
is determined by the associated eigenvalue difference
${\De_{JK}}$,
and the corresponding curve is the solution to the
equation $\De_{JK}L = \pi$
that fixes the occurrence of the first oscillation maximum.
An experiment is sensitive to oscillations of a particular type
if it lies in a region of the $\log L$-$\log E$ plane 
{\it above} the corresponding curve. 

In the standard three-neutrino massive model, 
the oscillations depend on $L/E$,
and the condition for the first oscillation maximum 
is described by a straight line with $\log L \propto \log E$ 
in the $\log L$-$\log E$ plane. 
In Lorentz-violating models,
a coefficient of the $a_L$ type
also produces a straight line for the 
the first oscillation maximum,
but with zero slope: $\log L$ is constant.
A coefficient of the $c_L$ type
produces a straight line with negative slope,
$\log L \propto - \log E$. 
Neither of the latter two behaviors is observed in the
neutrino-oscillation data.
However,
Lorentz-violating models with both types of coefficients
can produce complicated energy dependences
via seesaw effects
\cite{km1}. 
These effects produce the sensitivity curves for the tandem model
shown in Figs.\ \ref{fig:nusol}(a) and \ref{fig:antinusol}(a).

Note that the tandem model does exhibit
a limiting straight-line sensitivity behavior
with $\log L \propto \log E$ 
for $\De_{21}$
at atmospheric-neutrino energies and baseline distances. 
This feature is also present in the bicycle model,
and it matches the atmospheric-neutrino data
\cite{km2,mm}.
At high energies,
the limiting behavior of both 
$\De_{32}$ and $\De_{31} \equiv \De_{21}+\De_{32}$
is also linear, 
but with $\log L \propto -\log E$.
In the low-energy limit,
the sensitivity behaves 
as $\log L \propto -\log E$ for $\De_{21}$ and 
as $\log L \propto \log E$
for $\De_{32}$ and $\De_{31}$.

Figures \ref{fig:nusol}(b)-(d) and~\ref{fig:antinusol}(b)-(d)
show the energy dependence of the quantities 
$-4U_{Ja}U_{Jb}U_{Ka}U_{Kb}$ in Eq.\ \rf{tmosc}
that control the amplitude of the oscillation probabilities.
Unlike the standard three-neutrino massive model,
these quantities contribute to the energy dependence.
It turns out that they permit a description of the energy dependence 
in the solar-neutrino data
without resorting to matter-enhanced oscillations
\cite{msw}. 

\subsection{Comparison to data}

\begin{figure}
\centering
\includegraphics[width=\columnwidth]{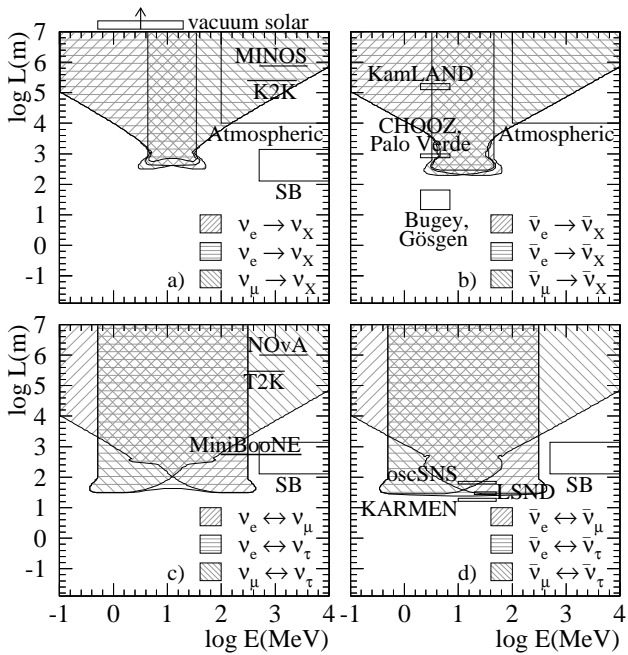}
\caption{
Oscillation probabilities in the $\log L$-$\log E$ plane for
(a) $\nu$ disappearance, 
(b) $\nub$ disappearance, 
(c) $\nu$ appearance, 
and 
(d) $\nub$ appearance.  
In (a) and (b),
the displayed regions have disappearance probability $P>10\%$.
Regions for disappearance of different flavors
are distinguished by different hatchings.  
In (c) and (d), 
the displayed regions have appearance probability $P>0.1\%$,
and the regions for different flavors
are again distinguished by different hatchings.  
Approximate $L$-$E$ sensitivity ranges 
for various oscillation experiments are displayed
as labeled rectangles.  
In (a), 
the rectangle for solar neutrinos lies well above the plot,
near $\log L \simeq 11$.
Short-baseline experiments with $E \gsim 1$ GeV are denoted SB.}
\label{fig:poscall}
\end{figure}

Some generic predictions of the tandem model 
for typical experimental sensitivities
are displayed in Fig.\ \ref{fig:poscall}.
These plots present information
about appearance and disappearance probabilities 
in the $\log L$-$\log E$ plane for both neutrinos and antineutrinos.
They provide an overview of the general predictions
of the model and relate them to some existing and future experiments.
Some details for the various classes of experiments
are given in the following subsections.

In Figs.\ 
\ref{fig:poscall}(a) and \ref{fig:poscall}(b),
regions above the curves 
correspond to disappearance probabilities greater than 10\%.
In Figs.\ \ref{fig:poscall}(c) and \ref{fig:poscall}(d),
regions above the curves correspond to appearance probabilities
greater than 0.1\%.
The figures also contain rectangles representing the regions
of sensitivity of particular experiments. 
The tandem model predicts a signal 
in an experiment for oscillations of a particular type 
when its rectangle overlaps 
the region above the oscillation curve.

\subsubsection{Solar}

In the standard three-neutrino massive model,
vacuum solar-neutrino oscillations show no energy dependence 
because $\De m^2_{\odot} \gg 10^{-10}$ eV$^2$
and the mixing matrix is energy independent.
In the tandem model,
however,
the long-baseline limit of vacuum oscillations 
exhibits an energy dependence 
arising from the mixing matrix $U_{Ja}$,
despite the loss of energy dependence in the factor
$\sin^2(\De_{JK}L/2)$.

\begin{figure}
\centering
\includegraphics[width=\columnwidth]{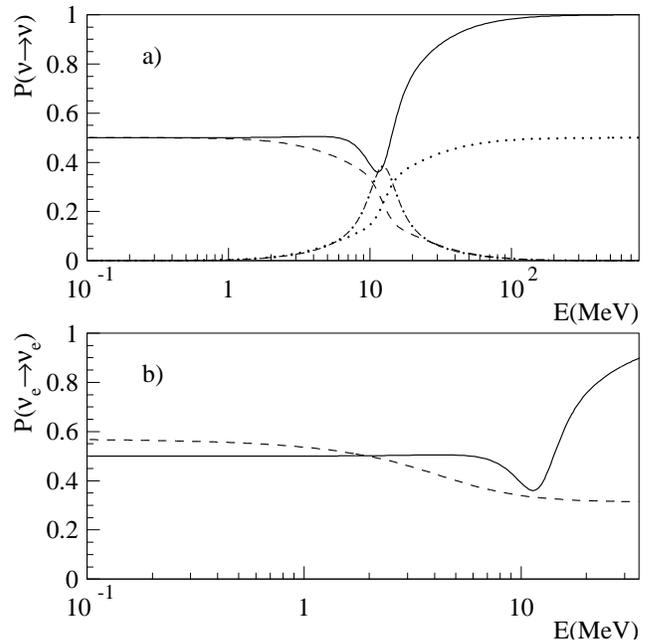}
\caption{Solar-neutrino oscillation probabilities
in the tandem model.  
(a) 
Curves representing survival probabilities for $\nu_e$ (solid),
$\nu_e\to\nu_\mu$ (dashed), 
$\nu_\mu\to\nu_\tau$ (dotted), 
and $\nu_\tau\to\nu_e$ (dash-dotted).  
(b) 
Survival probability of $\nu_e$ 
in the tandem model (solid line)
and in the standard three-neutrino massive model 
with a basic matter-induced effect (dashed). 
Note the different energy scales.
The effects of experimental position 
and energy resolution are not shown. 
}
\label{fig:solosc}
\end{figure}

The $\nu_e$ survival probability on the energy scale 
relevant to the solar neutrino problem is shown in 
Fig.\ \ref{fig:solosc}.
This model yields an oscillation probability of $50\%$ for 
pp neutrinos,
which have a continuous energy spectrum with endpoint at 0.420 MeV. 
The oscillation probability is also $50\%$ for $^7$Be neutrinos,
which generate two monoenergetic lines with $90\%$ at 0.862 MeV and 
$10\%$ at 0.384 MeV. 
For the $^8$B neutrinos,
which have a continuous energy spectrum with endpoint at 15.04 MeV,
the model yields a probability of $\simeq 40\%$.
These values are in reasonable agreement 
with the energy dependence observed
in the existing solar-oscillation data.

In the tandem model,
matter-enhanced oscillations 
\cite{msw}
play essentially no role in the description of solar neutrinos. 
In general,
matter-enhanced solar-neutrino oscillations can be viewed 
as arising from an effective coefficient for Lorentz violation
of the form 
\cite{km1}
$(a_{L,{\rm eff}})_{ee} = \sqrt{2} G_F n_e$,
where $n_e$ is the number density of electrons in the Sun.
An analytical approximation for $n_e$ 
\cite{ssm}
yields 
$(a_{L,{\rm eff}})_{ee} \simeq
1.87\times10^{-17}e^{-10.54R/R_\odot}$ MeV,
where $R$ is the distance from the core 
and $R_\odot$ is the solar radius.
In the tandem model,
substantial effects can arise only if 
$\cri E \lsim (a_{L,{\rm eff}})_{ee}$,
or $E \lsim 0.5e^{-10.54R/R_\odot}$ MeV.
For much of the solar-neutrino spectrum,
this energy is too small to allow appreciable effects
on the oscillation probability.
This result is confirmed by numerical calculations.

We remark in passing that an interesting feature 
of the tandem model in the 10 MeV region 
is a reduced suppression of the $\nub_e$ probability 
relative to that of $\nu_e$.  
This originates in CPT violation 
via the $\ari$ coefficient, 
and it may be relevant for $r$-process nucleosynthesis
following core-collapse in supernovae 
\cite{qfmmww}.

\subsubsection{Atmospheric}

\begin{figure}
\centering
\includegraphics[width=0.95\columnwidth]{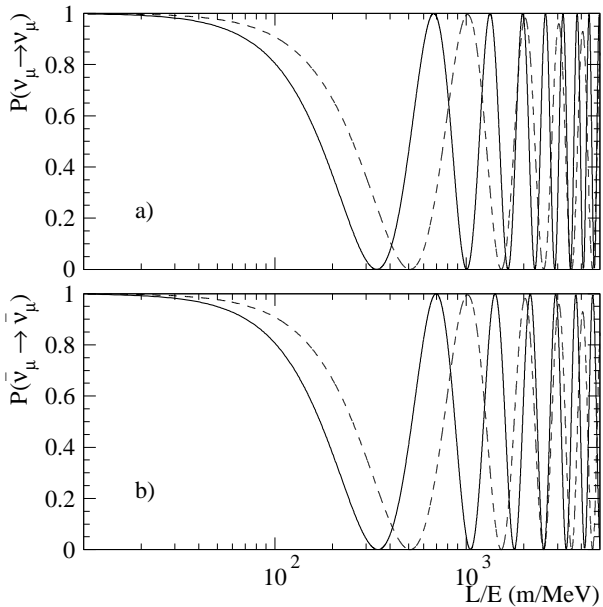}
\caption{
Survival probabilities for atmospheric neutrinos 
as a function of $L/E$
for (a) $\nu_\mu$ and (b) $\nub_\mu$.
Curves are shown for the tandem model (solid)
and for the standard three-neutrino massive model (dashed).
The effects of experimental position 
and energy resolution are not shown. 
}
\label{fig:atmosc}
\end{figure}

The atmospheric-neutrino results are well described 
by the tandem model.
At high neutrino energies, 
the effective tandem hamiltonian 
for neutrinos becomes approximately
\bea
h^\nu_{\mathrm{TM}}
& \approx & 
\left(
\begin{array}{ccc}
\cri E & ~ \ari  & ~ \ari \\
\ari & ~ 0 & ~ \ari \\
\ari  & ~ \ari & ~ 0
\end{array}
\right),
\eea
with a similar simplification for $h^{\bar\nu}_{\mathrm{TM}}$.
The $\cri$ and $\ari$ coefficients combine to produce
a Lorentz-violating seesaw mechanism
\cite{km1},
in which the model effectively reduces to a 
two-flavor limit with a pseudomass term
\cite{km2}.
The only significant probabilities are for
the transitions $\nu_\mu\mix\nu_\tau$ and $\nub_\mu\mix\nub_\tau$,
which involve $\De_{21}$ 
and are large only at energies above about 100 MeV. 
All amplitudes other than 
$U_{2 \mu}U_{2 \tau}U_{1 \mu}U_{1 \tau}$ 
approach zero at high energies.

A comparison of atmospheric-oscillation probabilities
in the tandem model 
with those in the standard three-neutrino massive model
is provided in Fig.\ \ref{fig:atmosc}.
In both models,
the disappearance probability is maximum 
in the region with $L/E =$ 300-600 m/MeV,
which corresponds to the results reported 
by the Super-Kamiokande experiment 
\cite{Super-Kosc}.
This feature is common to both $\nu_\mu$-$\nu_\tau$ 
and $\bar\nu_\mu$-$\bar\nu_\tau$ oscillations. 
The tandem model contains CPT violation,
but it is invisible in the high-energy region. 
These results are also consistent 
with the recent atmospheric measurement from the MINOS experiment 
\cite{MINOS}.

\subsubsection{Long-baseline reactor}

At low neutrino energies, 
the effective tandem hamiltonian for antineutrinos
becomes approximately
\bea
h^{\bar\nu}_{\mathrm{TM}}
& \approx & 
\left(
\begin{array}{ccc}
0  & ~ -\ari  & ~ -\ari \\
-\ari & ~ 0 & ~ -\ari \\
-\ari  & ~ -\ari & ~ \mrte 
\end{array}
\right).
\eea
The component $\mrte$ creates 
another Lorentz-violating seesaw
\cite{km1}.
Here,
it results in large $\nub_e\mix\nub_\mu$ oscillations 
involving $\De_{21}$,
with no contribution from other antineutrino flavors. 
The only oscillations are proportional to 
$U_{2 e}U_{2 \mu}U_{1 e}U_{1 \mu}$.

\begin{figure}
\centering
\includegraphics[width=0.85\columnwidth]{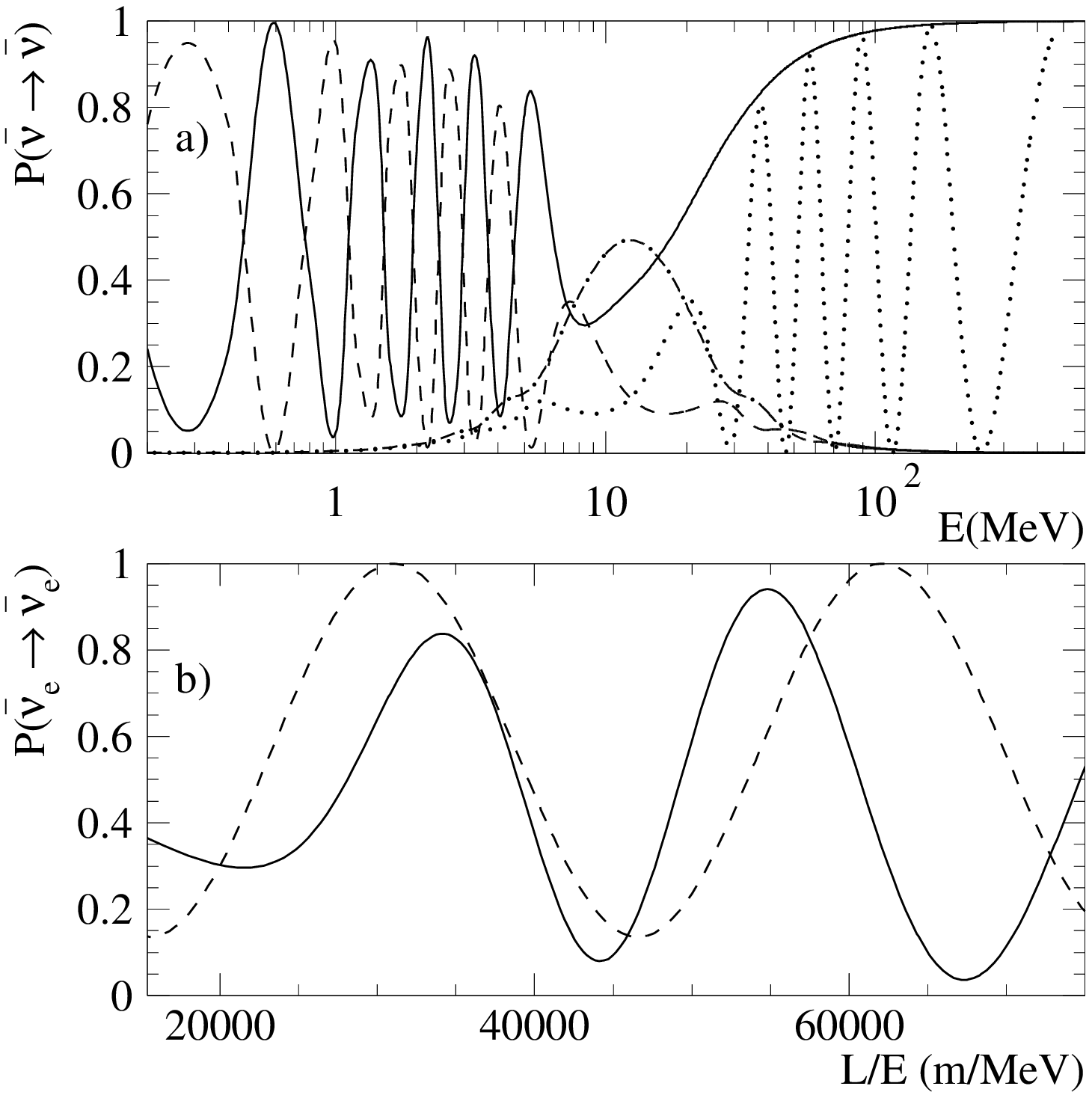}
\caption{Survival probabilities 
for long-baseline reactor antineutrinos.
(a) 
Survival probabilities as a function of $E$ for $\nub_e$ (solid),
$\nub_e\to\nub_\mu$ (dashed), 
$\nub_\mu\to\nub_\tau$ (dotted), and 
$\nub_\tau\to\nub_e$ (dash-dotted). 
(b) 
Survival probabilities for $\nub_e$ as a function of $L/E$
in the tandem model (solid)
and in the standard three-neutrino massive model (dashed). 
The effects of experimental position 
and energy resolution are not shown. 
}
\label{fig:kamosc}
\end{figure}

The oscillatory shape reported by the KamLAND experiment 
\cite{KamLANDosc}
is reproduced by the choice of values \rf{values}.
Figure \ref{fig:kamosc} shows the $\nub_e$ survival probability  
as a function of energy and $L/E$. 

Unlike the situation for atmospheric neutrinos, 
the CPT violation in the tandem model emerges  
in this $L$-$E$ range 
as a difference between $\nu_e$ and $\bar\nu_e$ 
survival probabilities.
However, 
this difference is unobservable in the KamLAND experiment 
because the sources produce only $\nub_e$.
Note that the $\nu_e\to\nu_\mu$ oscillation probability 
is substantial only at low energies. 
This means the tandem model predicts a null signal 
for future long-baseline $\nu_e$-appearance experiments 
such as NOvA \cite{NOvA} 
and T2K \cite{T2K}.

\subsubsection{Short-baseline reactor and accelerator}

\begin{table*}
\renewcommand{\arraystretch}{1.5}
\begin{tabular}{ccccccc}
\hline
\hline
Experiment \quad & \quad Ref.\ \quad 
& \quad Oscillation channel \qquad & \pt{x} Source \pt{x} \quad
& \quad \pt{x} Baseline \pt{x} \quad & \quad Energy \qquad 
& \pt{xx} Status \pt{xx} \qquad \\
\hline
Bugey & \cite{Bugey} & $\nub_e \to X$  & 
reactor  & 15,40~m & 3~MeV & null result \\ 
G\"{o}sgen & \cite{Gosgen} & $\nub_e \to X$  & 
reactor & 38, 46, 65~m & 3~MeV & null result \\ 
Palo Verde & \cite{PaloVerde} & $\nub_e \to X$  & 
reactor & 750, 890~m & 3~MeV & null result \\ 
CHOOZ & \cite{CHOOZ} & $\nub_e \to X$  & reactor & 
1~km & 3~MeV & null result \\
KARMEN & \cite{KARMEN} & $\nub_\mu \to \nub_e$ & 
accelerator & 18~m & 40~MeV & null result \\
LSND & \cite{LSND2} &  $\nub_\mu \to \nub_e$ &  
accelerator & 30~m & 40~MeV & signal \\
OscSNS & \cite{OscSNS} & $\nub_\mu \to \nub_e$ & 
accelerator & 60~m & 40~MeV & planning \\
MiniBooNE & \cite{MiniBooNE} & $\nu_\mu \to \nu_e$ & 
accelerator & 550~m & 800~MeV & ongoing \\
CDHS & \cite{CDHS} & $\nu_\mu \to X$ & accelerator & 
130~m & 1~GeV & null result \\
BNL-E776 & \cite{BNLE776} & $\nu_\mu \to \nu_e$, 
$\nub_\mu \to \nub_e$ & accelerator & 1~km & 1.4~GeV & null result \\
CHORUS & \cite{CHORUS} & $\nu_\mu \to \nu_\tau $ & 
accelerator & 600~m & 27~GeV & null result \\
NOMAD & \cite{NOMAD} & $\nu_\mu \to \nu_\tau$ & 
accelerator & 600~m & 45~GeV & null result \\
CCFR & \cite{CCFR} & $\nu_\mu \to \nu_e$, 
$\nub_\mu \to \nub_e$ & accelerator & 1~km & 140~GeV & null result \\
NuTeV & \cite{NuTeV} &$\nu_\mu \to \nu_e$, 
$\nub_\mu \to \nub_e$  & accelerator & 1~km & 150~GeV & null result \\
\hline
\hline
\end{tabular}
\caption{
Recent and near-future experiments
involving short-baseline oscillations 
at reactors and accelerators.
The notation $\nu_a \to \nu_b$ 
indicates an appearance search,
while $\nu_a \to X$ indicates a disappearance search.
}
\label{tb:sblexps}
\end{table*}

\begin{figure}
\centering
\includegraphics[width=\columnwidth]{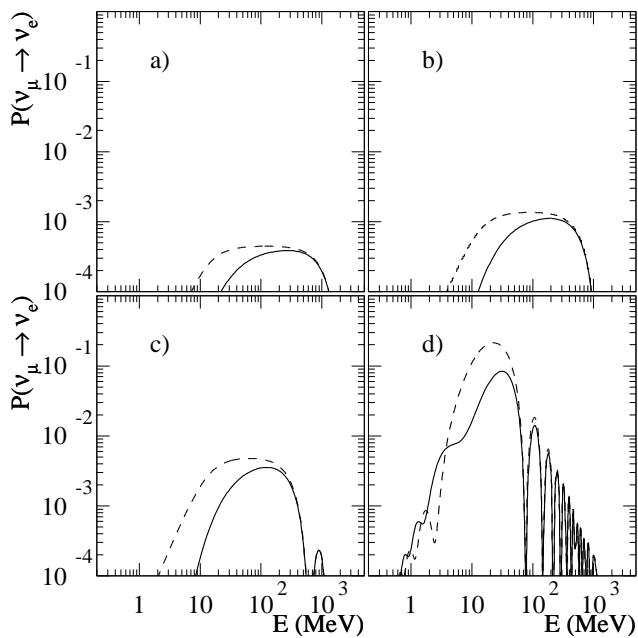}
\caption{
Oscillation probabilities as a function of $E$
for neutrinos (solid) and antineutrinos (dashed) 
in
(a) KARMEN,
(b) LSND, 
(c) the proposed OscSNS experiment,
and 
(d) the currently running MiniBooNE experiment.  
The effects of experimental position 
and energy resolution are not shown. 
} 
\label{fig:mbosc}
\end{figure}

In recent years, 
many experiments with short baselines 
\cite{KARMEN,LSND2,MiniBooNE,Bugey,Gosgen,%
OscSNS,PaloVerde,CHOOZ,CDHS,CHORUS,NOMAD,BNLE776,CCFR,NuTeV}
have searched for oscillations 
using neutrinos from reactor and accelerator sources. 
The results are summarized in 
Table \ref{tb:sblexps}.
The predictions of the tandem model
relevant at these baselines and energies
are discussed in this subsection.

In the tandem model,
oscillation probabilities at short baselines are typically small,
so they are undetectable in reactor-disappearance experiments 
with sensitivities of order 10\%.  
This can be seen from 
Fig.\ \ref{fig:poscall}.
Conceivably,
a sensitive reactor experiment with a baseline of $\gsim$ 1 km 
might observe a signal predicted by the tandem model
in the highest-energy bins of the $\nub_e$ spectrum.

For short baselines $\lsim$ 1 km,
oscillations of neutrinos at high energies $\gsim$ 100 MeV
are strongly suppressed,
as can also be seen from
Fig.\ \ref{fig:poscall}.
The oscillation behavior in the high-energy region is 
similar to that of atmospheric neutrinos, 
and any experiments with high-energy neutrinos
that are insensitive to atmospheric-parameter oscillations 
are also insensitive to oscillations in the tandem model.  

In contrast, 
for short-baseline low-energy accelerator experiments, 
the nonzero mass term 
in the tandem model
results in a small oscillation probability
that lies on the edge of experimental sensitivity.
The $\nu_\mu \to \nu_e$ and $\nub_\mu \to \nub_e$
oscillation probabilities 
for KARMEN, LSND, OscSNS, and MiniBooNE
are shown in Fig.\ \ref{fig:mbosc}.
The tandem model yields an oscillation probability 
on the order of 0.05-1\%
in the detectable energy range for all these experiments.

In particular,
the model yields an acceptable value 
for the oscillation probability 
of $(0.264 \pm 0.067 \pm 0.045) \%$
reported by LSND.
The oscillation probability is smaller for the KARMEN experiment 
due to the shorter baseline,
and it lies below the experimental sensitivity.
The tandem model predicts
substantial appearance signals for $\nu_e$ and $\nub_e$ 
in OscSNS and MiniBooNE.  
Note that these signals fall off rapidly with increasing energy.

\section{Summary}

In this work,
we have constructed a model of neutrino oscillations 
involving only three degrees of freedom
that describes all features of existing oscillation experiments,
including the LSND result. 
This tandem model uses renormalizable terms in the SME,
with one mass parameter and two coefficients for Lorentz violation.
The value of the mass parameter is compatible 
with a conventional seesaw mechanism,
and the coefficients for Lorentz violation
have Planck-suppressed magnitudes that 
are consistent with the measurements recently reported by LSND. 
The observed solar-neutrino suppression and energy dependence
are achieved without employing matter-enhanced oscillations. 

The tandem model makes quantitative predictions
that can be tested in the near future
with results from the MiniBooNE experiment 
and from various planned 
short- and long-baseline appearance experiments
such as OscSNS, NOvA, and T2K.
There may also be subsidiary signals
involving sidereal variations or other 
unconventional physics.
In any case,
the tandem model represents a candidate alternative to the
standard three-neutrino massive model,
and its existence reconfirms the key role 
of neutrino experiments in the ongoing exploration of physics
beyond the Standard Model. 

\section*{Acknowledgments}
\label{Acknowledgments}

This work is supported in part
by DOE grant DE-FG02-91ER40661, Tasks B and C,
and by NSF grants NSF-PHY-0100348 and NSF-PHY-0457219.

\appendix

\section{Diagonalization}

The general symmetric effective hamiltonian represented by 
the $3\times3$ matrix
\begin{eqnarray}
h_{\rm eff }
& = & 
\left(
\begin{array}{ccc}
\hee & \hem & \hte \\
\hem & \hmm & \hmt \\
\hte & \hmt & \htt
\end{array}
\right)
\end{eqnarray}
can be analytically diagonalized,
$h_{\rm eff } = U^\dagger E_{\rm eff} U$,
by solving the cubic eigenvalue equation
for the eigenvalues $E_J$ 
and then obtaining the diagonalizing matrix $U_{Ja}$.
In the effective hamiltonians \rf{tandemh} and \rf{antitandemh}
for the tandem model,
the entries are all real and given by 
$\hee = \cri E$,
$\hmm = 0$,
$\htt = \mrte$,
and 
$\hem = \hte = \hmt = \pm \ari$.

The first step is to solve the cubic eigenvalue equation.
This takes the form
\bea
\la^3 + a\la^2 + b\la + c = 0,
\label{cubic}
\eea
where 
\bea
a &=& -\tr{h_{\mathrm{eff}}},
\nonumber\\
b &=& 
\half [\tr{h_{\mathrm{eff}}}]^2
- \half \tr{h_{\mathrm{eff}}^2},
\nonumber\\
c &=& - \det{h_{\mathrm{eff}}}.
\eea
For the tandem model,
these quantities are
\bea
a &=& - \cri E - \mrte,
\nonumber\\
b &=& \cri \mrt/2 - 3 \ari^2,
\nonumber\\
c &=& \ari^2\left( \cri E \mp 2\ari + \mrte \right).
\eea
Note that $b$ is energy independent and 
that CPT violation contributes only to the determinant quantity $c$.

It is convenient to introduce the combinations of $a$, $b$, and $c$
given by 
\bea
Q &=& \frac{1}{9}(a^2-3b), 
\nonumber\\
R &=& \frac{1}{54}(2a^3-9ab+27c),
\nonumber\\
\th &=& \cos^{-1}\left( \fr{R}{\sqrt{Q^3}} \right). 
\eea
In terms of these combinations, 
the energy eigenvalues $E_J$ can be expressed as 
\bea
E_1 &=& -2\sqrt{Q}\cos\left(\fr{\th}{3} \right)-\frac{1}{3}a ,
\nonumber\\
E_2 &=& -2\sqrt{Q}\cos\left(\fr{\th +2\pi}{3} \right)-\frac{1}{3}a, 
\nonumber\\
E_3 &=& -2\sqrt{Q}\cos\left(\fr{\th -2\pi}{3} \right)-\frac{1}{3}a.
\eea

In general,
the solution to the cubic \rf{cubic}
involves three unequal real eigenvalues $E_J$
provided the discriminant $D\equiv R^2-Q^3$ is negative definite.
For the tandem model,
the values \rf{values} chosen for the model
satisfy this condition for all neutrino energies $E$.
With these values,
the discriminant $D$ increases with $E$ to a maximum near 10 MeV,
then decreases as $E$ increases. 
The appearance of the maximum around 10 MeV reflects the merger
of the two Lorentz-violating seesaw mechanisms near that energy.
Also,
since the eigenvalues $E_J$ of the tandem-model effective hamiltonian
are distinct for all energies $E$,
no two eigenvalue differences $\De_{JK}$ coincide exactly
for any given $E$.
This fact is reflected in 
Figs.\ \ref{fig:nusol}(a) and \ref{fig:antinusol}(a):
if two eigenvalue differences were to coincide exactly
then the logarithm of the third would diverge, 
a behavior absent from the figures.

The mixing matrix $U_{Ja}$ is found to take the form 
\bea
U &=& 
\left(
\begin{array}{ccc}
{B_1C_1}/{N_1} & ~{B_2C_2}/{N_2} & ~{B_3C_3}/{N_3} \\
{C_1A_1}/{N_1} & ~{C_2A_2}/{N_2} & ~{C_3A_3}/{N_3} \\
{A_1B_1}/{N_1} & ~{A_2B_2}/{N_2} & ~{A_3B_3}/{N_3}  
\end{array}
\right),
\eea
where
\bea
A_J &=& \hmt(\hee-E_J)-\hte\hem, 
\nonumber\\
B_J &=& \hte(\hmm-E_J)-\hem\hmt, 
\nonumber\\
C_J &=& \hem(\htt-E_J)-\hmt\hte,
\eea
and the normalization for each fixed $J$ is
\bea
N_J &=& \sqrt{A_J^2B_J^2+B_J^2C_J^2+C_J^2A_J^2}.
\eea

\end{document}